# Itinerant to localized transition of $f$ electrons in antiferromagnetic superconductor UPd$_2$Al$_3$

Shin-ichi Fujimori[1], Yuji Saitoh[1], Tetsuo Okane[1], Atsushi Fujimori[1,2], Hiroshi Yamagami[1,3], Yoshinori Haga[4], Etsuji Yamamoto[4] & Yoshichika Ōnuki[4,5]

[1]*Synchrotron Radiation Research Unit, Japan Atomic Energy Agency, Sayo, Hyogo 679-5148, Japan.*

[2]*Department of Complexity Science and Engineering, University of Tokyo, Kashiwa, Chiba 277-8561, Japan*

[3]*Department of Physics, Faculty of Science, Kyoto Sangyo University, Kyoto 603-8555, Japan*

[4]*Advanced Science Research Center, Japan Atomic Energy Agency, Tokai, Ibaraki 319-1195, Japan*

[5]*Graduate School of Science, Osaka University, Toyonaka, Osaka 560-0043, Japan*

**Electrons in solids have been conventionally classified as either band-like itinerant ones or atomic-like localized ones depending on their properties. For heavy Fermion (HF) compounds, however, the *f* electrons show both itinerant and localized behaviours depending on temperature[1]. Above the characteristic temperature $T^*$, which is typically of the order of few K to few tens K, their magnetic properties are well described by the ionic *f*-electron models, suggesting that the *f*-electrons behave as 'localized' electrons. On the other hand, well below $T^*$, their Fermi surfaces (FS's) have been observed by magneto-oscillatory techniques, and generally they can be explained well by the 'itinerant' *f*-electron model [2]. These two models assume totally different natures of *f*-electrons, and how they transform between localized and itinerant state as a function of temperatures has never been understood on the level of their electronic structures. Here we have studied the band structure of the HF antiferromagnetic superconductor $UPd_2Al_3$ well below and above $T^*$ by angle-resolved photoelectron spectroscopy (ARPES), and revealed the temperature dependence of the electronic structure. We have found that the *f*-bands, which form the FS's at low temperatures are excluded from FS's at high temperatures. The present results demonstrate how the same *f*-electrons show both itinerant and localized behaviours on the level of electronic structure, and provide an important information for the unified description of the localized and itinerant nature of HF compounds.**

UPd$_2$Al$_3$ shows an antiferromagnetic transition at $T_N$=14 K, and undergoes a transition into the superconducting state at $T$c=2 K[3]. One of the most remarkable features of this compound is the coexistence of the superconductivity and the magnetic ordering with a large magnetic moment (0.85$\mu_B$ per U atom). Its FS's at low temperatures were investigated by de Haas-van Alphen (dHvA) experiments[4], and they were well explained by the itinerant 5*f*-electron model as described by the band-structure calculations within the local-density approximation (LDA)[5]. At high temperatures, on the other hand, the local magnetic moment behaviour of the 5*f* electrons was also experimentally demonstrated. The magnetic susceptibility of UPd$_2$Al$_3$ follows the Curie-Weiss law above 50-70 K[6], suggesting almost completely localized magnetic moments at high temperatures. Therefore, 5*f*-electrons show both itinerant and localized properties depending on temperatures, and its crossover temperature *T*\* is about 50-70 K. There are two kinds of explanations for this temperature dependence of magnetic properties. One scenario is based on the conventional dense-Kondo picture where the localized *f*-moments at high temperatures are screened by a singlet coupling with the conduction electrons as decreasing the temperatures, and then condense into the itinerant heavy Fermi-liquid (HFL) states at further low temperatures. This scheme assumes that the same 5*f*-electrons are responsible for both the itinerant and localized properties. The other one is a two-component model, which assumes that the different subsystems of 5*f* electrons in one uranium atom site are responsible for each of the itinerant and localized properties[7]. In this scheme, it is assumed that one of the three U 5*f* electrons of one uranium atom is itinerant, while the remaining two electrons are localized. The former gives itinerant properties at low temperatures, while the latter explains the localized magnetic properties at high temperatures. Then, the temperature dependence of the magnetic susceptibility is explained by a crystalline electronic field model within the completely 'localized' U 5$f^2$ subsystem[3]. The model further conjectures that the unconventional superconductivity in UPd$_2$Al$_3$ originates from interaction between these two subsystems. Therefore, electronic structure of UPd$_2$Al$_3$ is still unresolved and one of the most controversial problems in heavy Fermion physics.

In recent years, ARPES experiments in the soft X-ray (SX) region (SX-ARPES) have been recognized as an useful technique to study the bulk electronic structure of strongly correlated electron materials[8]. One of the advantages of this technique is that the spectra are bulk-sensitive compared with conventional ARPES experiments with low-$h\nu$ photons. For uranium compounds, a further advantage is the large photoemission cross section of U 5$f$ states compared with other ligand $s$, $p$, and $d$ states in the soft X-ray region. Therefore, the SX-ARPES is an unique experimental method to obtain the bulk-sensitive and U 5$f$ dominant band structures as well as FS's of uranium compounds[9]. To elucidate how the U 5$f$ electrons of UPd$_2$Al$_3$ can show both 'itinerant' and 'localized' properties, we have performed ARPES experiments in the soft X-ray (SX) region (SX-ARPES) at temperatures well below and above $T^*$ (20 K and 100 K), and observed remarkable temperature dependence on its band structure.

We first show angle-integrated photoemission (AIPES) spectra measured in the SX region to see the energy distributions of the U 5$f$ states. In Fig. 1a, we show the AIPES spectra of UPd$_2$Al$_3$ measured with $h\nu$=400 and 800 eV at 20 K. In this photon energy range, the contributions from the U 5$f$ and Pd 4$d$ states are dominant, and those from the Al 3$s$, 3$p$ and U 6$d$ states are one to two order of magnitude smaller[10]. In Fig. 1a, we have also indicated the partial density of states (DOS) of the U 5$f$ and the Pd 4$d$ states obtained by relativistic band-structure calculation within the LDA[11]. Characteristic features of the measured spectra are well reproduced by the calculated DOS. Comparison with the calculation shows that the Pd 4$d$ states are distributed mainly below $E_B$ = 3 eV and the U 5$f$ states around the $E_F$. To extract contributions from the U 5$f$ states, we have subtracted the spectrum measured at $h\nu$=400 eV from that measured at $h\nu$=800 eV as shown in Fig. 2b since the photoemission cross section of the U 5$f$ states relative to that of the Pd 4$d$ states increases by a factor of two in going from $h\nu$=400 eV to $h\nu$=800 eV[10]. The spectrum measured at $h\nu$=400 eV has been broadened to simulate the energy resolution of the spectrum measured at $h\nu$=800 eV. Both spectra have been normalized so as to match the tail of Pd 4$d$ states located within $E_B$= 2-3 eV with each other. The difference spectrum representing the U 5$f$ distribution as indicated as the blue solid line in Fig 1b shows a sharp peak just below $E_F$, and a broad satellite structure extending to around $E_B$=1.5 eV. This 5$f$ DOS obtained by SX-PES is very similar to the one previously obtained by UV-PES by Takahashi *et al*[12] except that the relative intensity of this satellite structure to that of the quasi-particle bands is weaker in SX-PES than in UV-PES, and that the positions of the broad satellite structure are slightly different between SX-PES and UV-PES. These differences might be due to the higher bulk-sensitivity of SX-PES. In Fig. 2b, we compare the U 5$f$ difference spectrum (blue line) with the calculated U 5$f$ DOS broadened by the Gaussian function representing the instrumental resolution (yellow line). Since the experimental sharp peak structure located around $E_F$ corresponds to the calculated DOS, this peak should originate from the itinerant quasi-particle bands. The experimentally observed sharp peak just below $E_F$ is narrower than the calculated DOS, indicating that the 5$f$ bands are renormalized. On the other hand, the broad satellite structure on the deeper binding energy side ($E_B$>0.5 eV) cannot be explained by the band structure calculation, suggesting that it might originate from electron correlation effect. In the previous work [12], it was proposed that this broad structure is the contribution from the completely localized U 5$f^2$ subsystem. If this is the case, there should be only little energy dispersion as has been observed in the ARPES study on the localized U 5$f$ system UPd$_3$[13].

To examine the U 5$f$-derived band structure of UPd$_2$Al$_3$, we have performed SX-ARPES experiments at $h\nu$=595 eV. Before we show the ARPES spectra, we indicate the hexagonal Brillouin zone of UPd$_2$Al$_3$ in the paramagnetic phase in Fig. 2a. In Fig. 2b, we show the momentum position of the present ARPES cut by red curve. It can be understood that the present ARPES cut traces near the L-H-A high symmetry line. In Fig. 2c we show the density plot of ARPES spectra measured at 20 K. Some complicated dispersive features were clearly observed in the spectra, suggesting the existence of a well-defined cleavage surface. From the experimental energy distribution of U 5$f$ states, it is concluded that the narrow quasi-particle bands around $E_F$ were originated from U 5$f$ states. On the other hand, some strongly dispersive bands were clearly observed in the energy region of $E_B$=0.5-1.0 eV, where the broad satellite structure was observed in U 5$f$ AIPES spectra. Therefore, the origin of the broad satellite structure in AIPES spectra is the contribution not from less-dispersive localized U 5$f$ states but from the strongly dispersive bands with finite contribution from U 5$f$ states. In Fig. 2c we show the results of the energy band calculation for the L-H-A direction, in which all U 5$f$ electrons are treated as being itinerant. Contribution from the U 5$f$ and Pd 4$d$ states in each band are also indicated on the colour scale. Although the overall agreement between experiment and calculation is not so complete, some calculated bands have a correspondence to the experimental bands. Especially, bands with large contributions from Pd 4$d$ states located at $E_B$>1.0 eV (band 14 and part of band 15-16) have a good correspondence to ARPES spectra. In the near $E_F$ part of the spectra, quasi-particle band features, particularly the features around L point, have a good correspondence with band 18 although it is somewhat narrowed in the experiment. On the other hand, the contributions from U 5$f$ states are mostly distributed in $E_B$<0.5 eV in the band structure calculation while they are mixed with the bands located $E_B$=0.5-1.0 eV in the ARPES spectra. This is the most significant difference between the experiment and the band structure calculation, suggesting the importance of electron correlation effect in this compound.

We also measured the ARPES spectra at higher temperatures. In Fig. 3a, we show the comparison of the ARPES spectra of UPd$_2$Al$_3$ measured at 20K and 100 K, which were sufficiently lower and higher than $T^*$ (~50 K). Although the essential structures of the spectra were very similar between 20 K and 100 K, noticeable differences between them were observed. The quasi-particle peaks located at around $E_F$ show strongly momentum dependent changes. In going from 20 K to 100 K, the peak intensity is decreased at around L point while it is rather increased at around H point. Meanwhile, its position uniformly moves toward slightly higher binding energy sides as the temperatures is increased. The changes were observed at higher binding energy sides also. However, here we note that the changes in $E_B$>1.5 eV are mainly the changes of peak intensity, and their positions are not changed. This suggests that those changes are not originated from the band structures. Its origin is not clear at present, but it might be the changes of the inelastic scattering of photoelectrons due to the drastic changes of conductive properties of this compound. To check the validity of these temperature dependences of ARPES spectra, we have performed the different ARPES scans for different $UPd_2Al_3$ single crystal, and confirmed this temperature dependences of the ARPES spectra is reproducible. In addition, we have further measured the temperature dependence of ARPES spectra of $UNi_2Al_3$ whose characteristic temperature is higher than 300K[14]. In Fig 3b, we show the comparison of the APRES spectra of $UNi_2Al_3$ measured at 20 K and 100 K. It is clear that the spectra are not changed in these two temperatures. Therefore, we conclude that the temperature dependence of the ARPES spectra of $UPd_2Al_3$ at around $E_B=E_F-1.5$ eV are originated from the changes of its electronic structure below and above $T^*$.

We further analyze the temperature dependence of these ARPES spectra to understand the implication of these changes. In Fig. 4a and b, we show the ARPES spectra measured at 20 K and 100 K, which are divided by the Fermi-Dirac function convoluted with the Gaussian function to reveal the behaviours of the quasi-particle bands in the vicinity of $E_F$ more clearly. Now, it is clearly shown that the natures of the quasi-particle bands are very different between 20K (Fig. 4a) and 100 K (Fig. 4b). We have further taken the second derivatives of the ARPES spectra to identify the peak position in these spectra as shown in Fig. 4c (20 K) and 4d (100 K). In these figures, the bright part corresponds to the peaks in the ARPES spectra. At 20 K, the quasi-particle bands below $E_F$ disperse to above $E_F$, suggesting that they participate in the formation of the FS's. On the other hand, at 100 K, they move toward the higher binding energy side, and form less dispersive bands around $E_B=0.1$ eV. These changes imply that the quasi-particle bands of U 5$f$ origin at low temperatures, which form the HFL states, are excluded from the FS's at high temperatures. This behaviour of the quasi-particle bands is consistent with the optical conductivity measurement, where the renormalized Drude peak and the hybridization gap due to the formation of the heavy quasi-particle bands are observed at $T<50$ K while the usual metallic behaviour with less Drude weight is observed at $T>50$ K[15]. The usual metallic behaviour at $T>50$ K suggests that non-$f$ bands form FS at high temperatures. However, their contributions cannot be clearly observed in the present spectra since the lower energy part spectra are dominated by strong U 5$f$ contribution. These changes are consistent with the dense-Kondo scenario, where the heavy quasi-particle bands are formed by hybridization between the flat renormalized $f$-level located around $E_F$ and dispersive non-$f$ bands[16]. On the other hand, the temperature-induced changes in the deeper binding energy region ($E_B$<1 eV) were also clearly recognized in these figures. For example, at around the circles designated as 1 in Fig 4c and d, the structure located at $E_B$=1 eV moves toward $E_B$=1.1 eV in going from 20 K to 100 K. In addition, the complicated band structures indicated in the circles designated as 2 also have significant temperature dependences. Thus, the transition is accompanied with the drastic changes of the band structure of an energy order of 1 eV (~1000 K), which has not been expected in the theoretical prediction of periodic Anderson model[17]. Therefore, global nature of U 5$f$ electrons of UPd$_2$Al$_3$ including temperature dependence of band structures cannot be explained by any present theoretical frameworks, and the present results claim a new scheme to describe the itinerant and localized nature of $f$ electrons as well as the coexistence of magnetism and superconductivity in heavy fermion compounds.

## Methods

**Photoemission experiment and data analysis.** Photoemission experiments were performed at the soft X-ray beamline BL23SU of SPring-8[18] using a photoemission spectrometer equipped with a Gammadata-Scienta SES-2002 electron analyser. The AIPES spectra were measured by the angle-integrated mode of the spectrometer. The energy resolutions were 160 meV ($h\nu$=800 eV) and 80 meV ($h\nu$=400 eV). The energy and angular resolution of SX-ARPES experiments were set to 120 meV and 0.16° (corresponding to 0.036 Å$^{-1}$) respectively for photon energy of 595 eV to obtain a reasonable count rate. Samples were cooled using a closed-cycle He refrigerator. Sample temperature was measured using a chromel-AuFe thermocouple mounted close to the sample. The base pressure of the spectrometer was better than $2\times10^{-8}$ Pa. The sample orientation was measured *ex situ* using Laue photography. The position of the Fermi level was determined by the position of that of *in situ* evaporated gold film. The measured momentum positions were determined with a free-electron final-state model by taking the photon momentum into consideration. The momentum of the electron perpendicular to the surface with free-electron final state model is given by.

$$k_\perp = \sqrt{\frac{2m}{\hbar^2}\left(E_{\text{kin}}\cos^2\theta + V_0\right)} - k_{\perp\text{photon}} \ ,$$

where $E_{\text{kin}}$ is the kinetic energy of the photoelectron, $V_0$ is the inner potential, $\theta$ is the emission angle of photoelectron relative to the surface normal, and $k_{\perp\text{photon}}$ is the momentum of incident photon perpendicular to the surface respectively. We assumed that the inner potential was $V_0=12$ eV, which is a typical value for HF compounds[19]. The image plots shown in Fig's 4a and b are obtained by taking the second derivatives of the ARPES spectra. Overall, the second derivatives of the spectra have been used to reveal positions of the band structure. In the present procedure, we have added the second derivative of energy distribution curves, which is sensitive to weakly dispersive bands, and that of momentum distribution curves, which is sensitive to strongly dispersive bands, with an appropriate weight to make the band images most clearly visible. We applied the same procedures with same parameters for the ARPES spectra measured at 20 K and 100 K.

**Sample preparation.** Single crystals were characterized by the single crystal x-ray diffraction. Lattice parameters at room temperature were determined to be $a=$ 5.368 and $c=$4.189 Å, in good agreement with previous studies[20]. There is no defect at either Pd or Si sites within experimental accuracy of 1 %. The direction of the crystal axes were determined by the Laue method. Clean sample surface was obtained by cleaving the sample *in situ*.

**Acknowledgement**

We would like to acknowledge N. K. Sato, K. Miyake, N. Aso, G. Zwicknagl, G. H. Lander and A. Chainani for stimulating discussion and comments. The present work was financially supported by a Grant-in-Aid for Scientific Research from the Ministry of Education, Culture, Sports, Science, and Technology Japan under contact No.15740226, and a REIMEI Research Resources from Japan Atomic Energy Agency.

Correspondence and requests for materials should be addressed to S.-I.F. (e-mail: fujimori@spring8.or.jp).

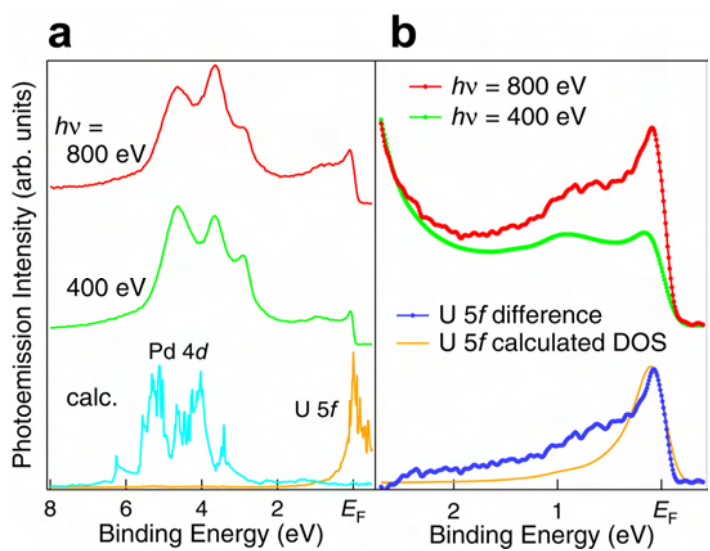

**Figure 1|Angle-integrated photoemission spectra of UPd$_2$Al$_3$. a**, The spectra measured at $h\nu$=400 and 800 eV, together with the calculated Pd 4$d$ and U 5$f$ density of states. **b**, The procedure of deriving of the experimental U 5$f$ partial density of states. The spectra measured at $h\nu$=400 eV (green curve) has been subtracted from that measured at $h\nu$=800 eV (red curve). The obtained U 5$f$ partial density of states is shown as blue solid curve and is compared with the calculated U 5$f$ partial DOS.

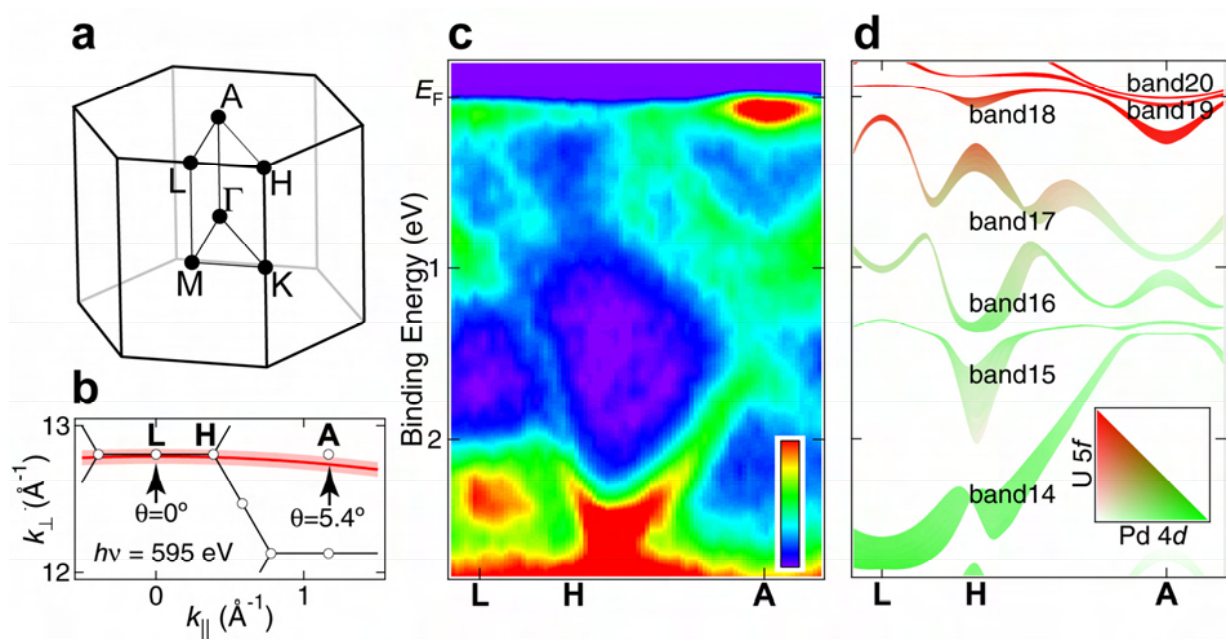

**Figure 2|SX-ARPES spectra of UPd$_2$Al$_3$. a**, the hexagonal Brillouin zone of UPd$_2$Al$_3$ in a paramagnetic phase. **b**, The position of the ARPES scan. The red curve represents the momentum position of ARPES cut with $h\nu$=595 eV. The shaded area indicates the momentum broadening for $k_\perp$ direction due to the finite electron escape depth. **c**, The density plot of the SX-ARPES spectra measured 20 K. **d**, The calculated energy dispersions to be compared with the experiment.

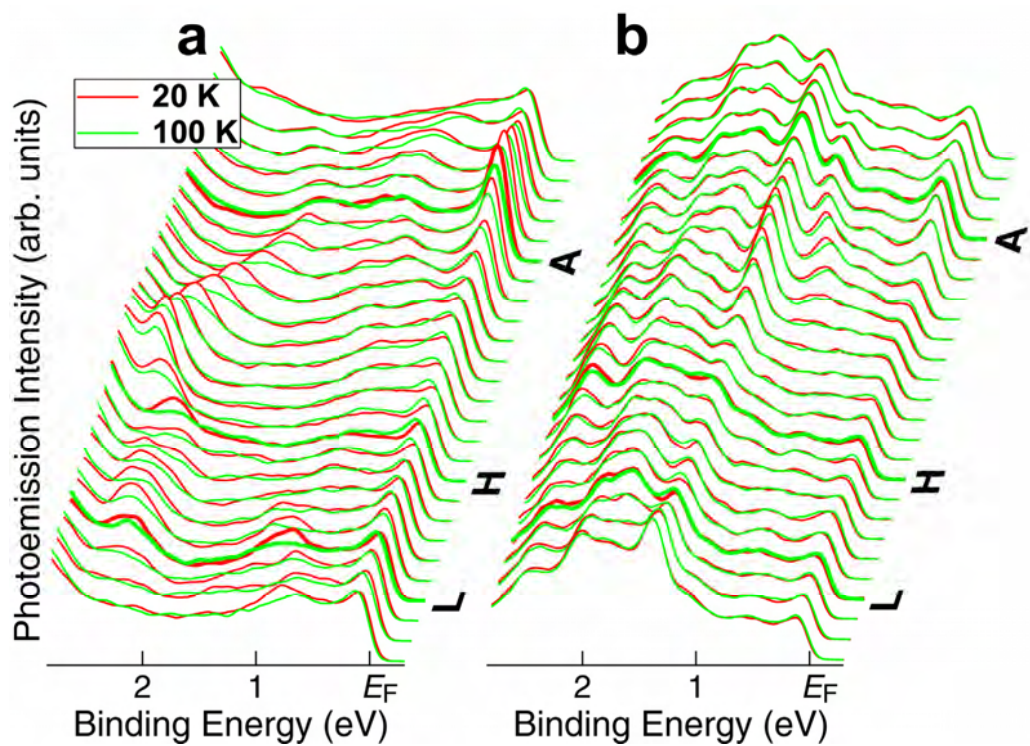

**Figure 3|Temperature dependence of ARPES spectra. a**, ARPES spectra of UPd$_2$Al$_3$ measured at 20 K and 100 K. **b**, ARPES spectra of UNi$_2$Al$_3$ measured at 20 K and 100K. The temperature dependences were observed in UPd$_2$Al$_3$ while they were not observed in UNi$_2$Al$_3$.

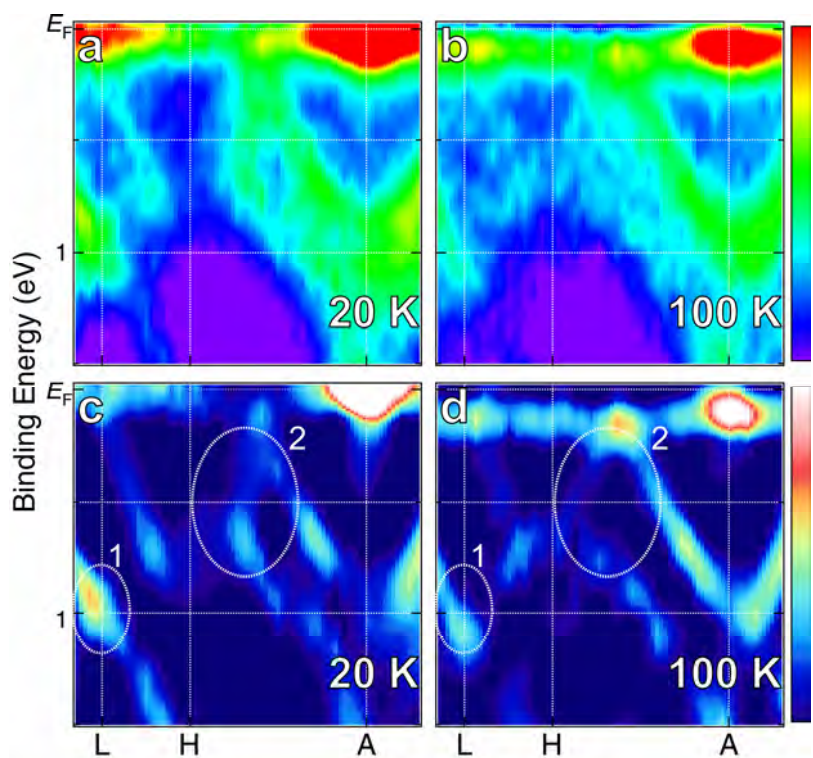

**Figure 4|Experimental band structures of UPd$_2$Al$_3$. a-b**, ARPES spectra divided by the convoluted Fermi-Dirac function. **c-d**, Experimental band structure derived by adding the second derivatives of EDC's and MDC's in a proper ratio. Higher intensity part corresponds to the peak position in the ARPES spectra. The circles indicate the positions of changes of the band structure in higher binding energy sides.